\documentclass {aa}
\usepackage{graphicx}
\begin{document}
                
\title{Are rotating strange quark stars good sources of gravitational waves?}


\author{Dorota Gondek-Rosi{\'n}ska\inst{1,2}
\and Eric Gourgoulhon\inst{1}
\and Pawe\l \ Haensel\inst{2,1}
}

%
\institute{Laboratoire de l'Univers et de ses Th\'eories, UMR 8102 du
C.N.R.S., Observatoire de Paris, F-92195 Meudon Cedex, France
\and Nicolaus Copernicus Astronomical Center, Bartycka 18, 00-716 Warszawa,
 Poland}

\offprints{D. Gondek-Rosi{\'nska}, \email{Dorota.Gondek@obspm.fr}}

\date{} \abstract{We study the viscosity
  driven bar mode (Jacobi-like) instability of rapidly rotating quark
  matter stars (strange stars) in general relativity. A triaxial,
  ``bar shaped'' compact star could be an efficient source of
  continuous wave gravitational radiation in the frequency range of
  the forthcoming interferometric detectors. We locate the secular
  instability point along several constant baryon mass sequences of
  uniformly rotating strange stars described by the MIT bag model.
  Contrary to neutron stars, strange stars with $T/|W|$ (the ratio of
  the rotational kinetic energy to the absolute value of the
  gravitational potential energy) much lower than the corresponding
  value for the mass-shed limit can be secularly unstable to  bar
  mode formation if shear viscosity is high enough to damp out any
  deviation from uniform rotation. The instability develops for a
  broad range of gravitational masses and rotational frequencies of
  strange quark stars. It imposes strong constraints on the lower limit of
  the frequency at the innermost stable circular orbit around rapidly
  rotating strange stars. The above results are robust for all linear
  self-bound equations of state assuming the growth time of the
  instability is faster than the damping timescale.  Whether the
  instability can grow or not  depends on many different physical
  quantities (e.g. value of viscosities (rather uncertain)). We discuss
  astrophysical scenarios where triaxial instabilities (r-mode and
  viscosity driven instability) could be relevant (a new born star,
  an old star spinning up by accretion) in strange stars described by
  the standard MIT bag model of normal quark matter. The
  spin evolution of a strange star strongly depends on the strange
  quark mass. Taking into account actual values of viscosities in
  strange quark matter and neglecting the magnetic field we show that
  Jacobi-like instability cannot develop in any astrophysicaly
  relevant temperature windows. The main result is that strange quark
  stars described by the MIT bag model can be accelerated to very high
  frequency in Low Mass X-ray binaries if the strange quark mass is
  consistent with values based on particle data tables.
 }

\maketitle

\section{Introduction}

The first generation laser interferometric gravitational wave
detectors such as LIGO, VIRGO and GEO600 should be fully operationally
soon and second generation detectors like LIGO II within a few
years. Triaxial instabilities of rotating compact stars (neutron
stars, strange quark stars) can play an important role as emission
mechanisms of gravitational waves in the frequency range of these
detectors (see e.g. Stergioulas 1998, 2003 or Andersson 2003 for a review).
A rapidly rotating relativistic star can spontaneously break its axial
symmetry if the ratio of the rotational kinetic energy to the absolute
value of the gravitational potential energy $T/|W|$ exceeds some
critical value (Gondek-Rosi\'nska \& Gourgoulhon 2002, Di Girolamo \&
Vietri 2002, Shapiro \& Zane 1998, Bonazzola et al. 1996, 1998).  The
$m=2$, $l=2$ ``bar mode'' instability is the fastest-growing
instability if a rotation rate is sufficiently high.

Neutron stars can be forced to rotate rigidly if viscosity is high
enough to damp out any deviation from uniform rotation.  The
viscosity-driven bar mode (Jacobi-like) instability of neutron stars
has been studied in general relativity by Bonazzola, Frieben \&
Gourgoulhon (1996, 1998), who find that only the stiffest equations of
state (EOS) allow symmetry breaking. Since for ``realistic'' EOS of
neutron stars the ratio of $T/|W|$ has a low value (usually lower than
0.14), this instability takes place only when a star is rotating at a
frequency close to the mass-shed limit and has high mass.

Strange quark stars (SS) are currently considered as a possible
alternative to neutron stars as compact objects (see e.g. Weber 1999 
for a review) . Claims of discovery have been recently announced (Drake et
al. 2002), but the evidence is not yet decisive (Walter \& Lattimer
2002).  It has been suggested by Gourgoulhon et
al. (1999), Gondek-Rosi\'nska et al. (2000a, b; 2001a) that a triaxial
instability could develop more easily in a uniformly rotating strange
star than in a neutron star. It was shown that independent of any
details of the EOS of strange matter, the ratio $T/|W|$ for self-bound
objects like strange stars can  be  even more than twice as high as for
any models of neutron stars. No stability analysis in general
relativity has been performed for strange stars untill now, because
numerical constraints prevented investigators from treating stars with strongly
discontinuous density profiles at the surface.  In the present paper we
perform a numerical study of the secular triaxial instability of rigidly
rotating strange stars in general relativity. We use the code
developed recently by Gondek-Rosi\'nska \& Gourgoulhon (2002) for
locating the secular bar mode instability points of uniformly rotating
incompressible fluid bodies in general relativity. Numerous tests have
been performed to assess the validity of the method and the accuracy
of the numerical code (Gondek-Rosi\'nska \& Gourgoulhon 2002).

In this paper we shall focus on the following questions: for
which gravitational masses and rotational frequencies of strange stars
does the Jacobi-like instability occur? Does gravitational radiation
set strong limits on spin rate of strange stars and on the frequency
at the innermost stable circular orbit around rapidly rotating strange
stars? What astrophysical scenarios does the viscosity driven
instability allow one to develop? How do our conclusions depend on physical
parameters of the MIT bag model?

The plan of the paper is as follows. The description of previous
studies of the bar mode instabilities is presented in Sec. 2. Section
3 and 4 are devoted to the description of the strange matter EOS used
by us and of the numerical technique we employed, respectively.  The
numerical results and discussion are presented in two sections Sec. 5
and 6. In Sec. 5 we examine the necessary condition for
the onset of the instability: that the frequency of a corotating mode
goes through zero in a frame co-rotating with the star. This condition
yields the critical angular velocity (at a given compactness) for
which the instability could start to operate, provided it is not
damped by other mechanisms.  The results are robust for any linear
self-bound EOS.  In Sec 6. we discuss the astrophysical scenarios when
triaxial instabilities could be relevant. This part includes an
estimate of different timescales (uncertain).  Finally, Sec. 7
provides a summary of our work.
 
\section{Previous studies of the bar mode instability}

Rapidly rotating compact stars may be subject to different kinds of
bar mode non-axisymmetric rotational instabilities.  

There exist two
different classes and corresponding time-scales for non-axisymmetric
instabilities: a dynamical instability, growing on the hydrodynamical
time-scale and secular instabilities which grow on much longer
dissipation time-scales.  In Newtonian theory, a rapidly rotating
incompressible fluid body (a Maclaurin spheroid) is secularly unstable
to a bar-mode formation, if $T/|W| > 0.1375$ (Jacobi/Dedekind
bifurcation point) and dynamically unstable if $T/|W| > 0.2738$
(Chandrasekhar 1969).  A secular instability can grow only in the
presence of dissipative mechanisms, such as {\it shear viscosity}
(Roberts \& Stewartson 1963) or {\it gravitational radiation} (the
Chandrasekhar-Friedman-Schutz instability (hereafter CFS instability;
Chandrasekhar 1970; Friedman \& Schutz 1978; Friedman 1978). Viscosity
dissipates mechanical energy but preserves angular momentum. As shown
by Christodoulou et al. (1995), the viscosity driven (Jacobi-like) bar
mode instability appears only if the fluid circulation is not
conserved. If on the contrary, the circulation is conserved (as in
inviscid fluids submitted only to potential forces), but not the
angular momentum, it is the gravitational radiation (Dedekind-like)
instability which develops instead.  If both viscosity and
gravitational radiation are operative, they act against each other.
Which of the two dissipative mechanisms dominates is highly dependent
on the viscosities and the temperature of a star.

The few numerical studies of secular instabilities in rapidly rotating
stars in general relativity show that non-axisymmetric modes driven
unstable by viscosity no longer coincide with those made unstable by
gravitational radiation. As was shown by recent post-Newtonian
(Shapiro \& Zane 1998, Di Girolamo \& Vietri 2002) and
relativistic (Gondek-Rosi\'nska \& Gourgoulhon 2002) studies of
viscosity-driven instability in rigidly rotating incompressible fluid
bodies, general relativity weakens this bar mode instability: the
values of $T/|W|$, eccentricity and rotation rate increase at the
onset of instability as the compaction parameter $GM/(R_{\rm circ}c^2)$
increases.  This stabilizing tendency of general relativity is in
agreement with previous studies of Jacobi-like instability of rigidly
rotating compressible (neutron) stars (Bonazzola, Frieben \&
Gourgoulhon 1996, 1998). The stabilizing effect of relativity is not
very strong, the critical value of $T/|W|$ for relativistic neutron
star models is ~$0.14$ (the Newtonian value is 0.1275 (Bonazzola et
al. 1996) for compressible bodies) and ~0.18 for very relativistic
(with proper compaction parameter $M/R=0.25$) incompressible fluid
body.

  On the other hand, the gravitational-radiation driven instability is
 strengthened by relativity. As was first proved by Friedman \& Schutz
 (1978) the Dedekind-like instability is generic in rotating stars -
 one can always find a mode that is unstable for a star with a given
 rotation rate. The most important pulsation modes for the CFS
 instability are the r-mode and f-mode (Stergioulas 1998, 2003,
 Andersson 2003, Andersson \& Kokkotas 2001, Stergioulas \& Friedman
 1998, Friedman and Lockitch 2002, for recent reviews of these
 instabilities).  For a wide range of realistic EOS, Morsink,
 Stergioulas \& Blattnig (1999) find that the critical $T/|W|$ for the
 onset of the instability in the $l=2$ f-mode is only $T/|W| \sim
 0.08$ for models with mass $1.4 M_\odot$ and $T/|W| \sim 0.06$ for
 maximum mass models.  However, based on the relative strength of the
 f-mode and r-mode instabilities for NS, one can expect that the
 r-mode instability will occur at lower rotation rates and in a larger
 temperature window than the f-mode instability, unless some damping
 mechanism favors a stronger damping at the r-mode frequency. The
 onset of the r-mode instability for SS has been studied by Madsen
 (2000) and  Andersson, Jones \& Kokkotas (2002).

 The dynamical instability against bar-mode deformation of rapidly
rotating polytropic stars was studied by Shibata, Baumgarte \& Shapiro
(2000) in full general relativity and by Saijo et al. (2001) in the
first post-Newtonian approximation. The authors considered
differentially rotating proto-neutron stars. Their main result is
that in general relativity the onset of dynamical instability occurs
for a somewhat smaller critical value of $(T/|W|)_{\rm crit}\sim
0.24-0.25$ than the Newtonian value $\sim 0.27$ for incompressible
Maclaurin spheroids.

\section{Equation of state}
The possibility of the existence of quark matter dates back to the
early seventies. Bodmer (1971) remarked that matter consisting of
deconfined up, down and strange quarks could be the absolute ground
state of matter at zero pressure and temperature.  If this is true
then objects made of such matter, so-called "strange stars", could exist
(Witten 1984). Typically, strange stars are modeled with an equation
of state  based on the phenomenological MIT-bag model of quark
matter (Alcock et al., 1986; Haensel et al., 1986), in which quark
confinement is described by an energy term proportional to the volume
(Farhi \& Jaffe 1984).

In the framework of this model the quark matter is composed of
massless u, d quarks, massive s quarks and electrons. There are three
physical quantities entering the model: the mass of the strange
quarks, $m_{\rm s}$, the bag constant, $B$, and the strength of the
QCD coupling constant $\alpha$.

The EOS of strange quark matter (SQM) is calculated using a
perturbation expansion in $\alpha$.  The expansion is treated in the
renormalization scheme. In particular, one has to use {\it
renormalized} values of $m_s$ and $\alpha$, which both depend on the
selected value of the renormalization point, $\rho_{\rm R}$ (see,
e.g., Farhi \& Jaffe 1984). The value of $\rho_{\rm R}$ is of the
order of typical energies in the quark-system under consideration.
Following Farhi \& Jaffe (1984), we will use $\rho_{\rm R}=313~$MeV.
\subsection{The strange quark mass}
The s quark mass given in the tables compiled by the Particle Group is 
the so called {\it current quark mass}, $m^{\rm (c)}_s=75-150~$MeV 
(Groom et al. 2000). Such a mass can be used in the systems in which the 
energies of the relative quark-quark motion $\gg 1$~GeV, say 100 GeV, 
which is many 
orders of magnitude larger than typical energies in SQM. Therefore, 
the value of $m_s$ has to renormalized using the renormalization group
equation, and its value calculated at $\rho_{\rm R}$ (Farhi \& Jaffe 1984). 
The lowest-order (in $\alpha$) renormalization group equation for $m_s$
indicates that renormalization will significantly increase the s-quark 
mass, and that with our choice of $\rho_{\rm R}$ we will have 
$m_s\sim 2 m_s^{\rm (c)}$. In what follows, we will use $m_s c^2 =
200~$MeV as a basic (standard) value of $m_s$ in SQM, but we will 
also present some results for $m_s=100$~MeV, to visualize sensitivity 
of our results to the value of $m_s$. 
\subsection{Models of EOSs of strange matter}

Our basic EOS corresponds to the standard values of the 
MIT bag model parameters of SQM:
\begin{itemize}
\item{} SQS1 - the standard MIT bag model: $m_{\rm s}c^2=200\ {\rm MeV}$,
$\alpha=0.2$, $B=56~{\rm MeV/fm^3}$;
\end{itemize}

 We shall also find the onset of the viscosity driven instability for
two other MIT bag models to check how our results depend on model
parameters:
\begin{itemize}
\item{} SQS0 - the simplified MIT bag model: $m_{\rm s}=0$, $\alpha=0$;
$B=60\ {\rm MeV/fm^3}$;
\item{} SQS2 - the ''extreme'' MIT bag model (relatively low strange
quark mass and $B$ but very high $\alpha$) : $m_{\rm s}c^2=100\ {\rm MeV}$,
$\alpha=0.6$, $B=40\ {\rm MeV/fm^3}$.
\end{itemize}

\section{Numerical codes}
\subsection{Axisymmetric equilibrium stars}

We have calculated axisymmetric models of rotating strange stars using
two different highly accurate relativistic codes.  The first code
developed by Gourgoulhon et al. (1999) is based on the multi-domain
spectral method of Bonazzola et al. (1998).  This code has been used
previously for calculating properties of rapidly rotating strange
stars described by the MIT bag model (Gourgoulhon et al. 1999,
Gondek-Rosi{\'n}ska et al. 2000a, 2001a, Zdunik et al. 2000a) and by
the Dey et al. (1998) model of strange matter (Gondek-Rosi{\'n}ska et
al. 2000b, 2001b). To compute very rapidly rotating axisymmetric
stationary configurations (when the ratio of polar to equatorial
radius is lower than $0.3-0.5$ depending on the 
compaction parameter - see
Sec.~III.A of Gondek-Rosi{\'n}ska \& Gourgoulhon 2002 for a
discussion), where the first code has difficulties converging to a
solution, we employ a second numerical code (Stergioulas \& Friedman
1995, see Stergioulas 1998 for a description). In this code the
equilibrium models are obtained following the KEH method (Komatsu et
al., 1989), in which the field equations are converted to integral
equations using appropriate Green functions.  Models computed by
both codes agree very well in all computed properties (Stergioulas,
Klu\'zniak \& Bulik 1999).

\subsection{The secular bar mode instability}

Recently Gondek-Rosi\'nska \& Gourgoulhon (2002) used multi-domain
method for calculating Jacobi-like instability of relativistic
incompressible fluid bodies. They have introduced surface-fitted
coordinates, which enable them to treat the strongly discontinuous
density profile at the surface of incompressible bodies. This avoids
any Gibbs-like phenomenon and results in a very high precision, as
demonstrated by comparison with the analytical result for the
Newtonian Maclaurin-Jacobi bifurcation point, that their code has
retrieved with a relative error of $10^{-6}$.  
Their  results are in
close agreement with the very recent post-Newtonian calculations of
viscosity driven bar mode instability in rigidly rotating
incompressible fluid bodies by Di Girolamo \& Vietri (2002) (the
relative difference is less than $1\%
$.)

The method of finding the secular bar mode instability point presented
there is an improvement on an earlier method by Bonazzola, Frieben \&
Gourgoulhon (1996, 1998). The approach consists of perturbing an
axisymmetric and stationary configuration and studying its evolution
by constructing a series of triaxial quasi-equilibrium configurations.
If the perturbation is damped (resp. grows) during the interation,
the equilibrium configuration is declared to be stable
(resp. unstable) (see Gondek-Rosi\'nska \& Gourgoulhon 2002 for a
description).  The above method does not require us  to specify some value
of viscosity. Whatever this value, the effect of viscosity is
simulated by the rigid rotation profile that is imposed at each step
in the iterative procedure. If one considers the iteration as mimicking
some time evolution, this means that the time elapsed between two
successive steps has been long enough for the actual viscosity to have
rigidified the fluid flow. Consequently, using this method one cannot get 
the instability time scale, which depends upon the actual value
of the viscosity, but only  the instability.

We use the same code as Gondek-Rosi\'nska \& Gourgoulhon (2002) for
finding the viscosity driven instability points of rapidly rotating
strange stars described by the MIT bag model.

\section{Calculations and results} \label{s:results}

\subsection{Effect of relativity on the secular bar mode instability} 
\begin{figure}
\includegraphics[width=0.9\columnwidth]{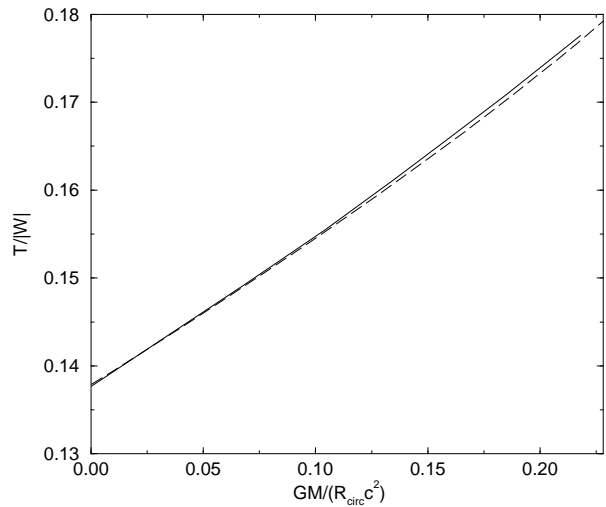}
\caption{
Effect of general relativity 
on the critical value of the kinetic
energy to the absolute value of the gravitational potential energy for
different models of strange quark stars described by MIT bag model
(solid line) and incompressible fluid stars calculated by
Gondek-Rosi\'nska \& Gourgoulhon (2002) (dashed line). The ratio of
$T/|W|$ for compaction parameter $GM/(R_{\rm circ} c^2) =0$
corresponds to the classical Newtonian Jacobi/Dedekind bifurcation
point, 0.1375.
}
\label{fig:TWMR}
\end{figure}

\begin{figure}
\includegraphics[width=0.9\columnwidth]{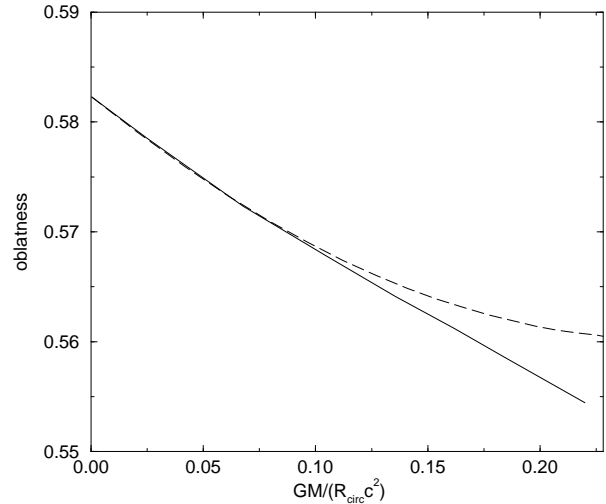}
\caption{Critical values of stellar oblateness as a function of the
proper compaction parameter for different models of strange quark
stars described by MIT bag model (solid line) and incompressible fluid
stars (dashed line) }
\label{fig:oblat}
\end{figure}
In this section we are focusing on the effect of relativity on the
Jacobi-like instability in the case of strange quark stars. Strange
stars in contrast to ordinary NS have a stiff EOS at the surface and
are soft inside and their density profiles are flat. Rotating strange
stars with small masses, lower than $0.1\ M_\odot$, can be described
very well by a uniform-density fluid, i.e., the classical Maclaurin
spheroids (see Amsterdamski et al. 2002 for comparison between fully
relativistic calculations of rapidly rotating low mass strange stars
and Maclaurin spheroids).  Widely used indicators for the onset of
viscosity driven instability are {\it the ratio of the kinetic energy
  to the absolute value of the gravitational potential energy} $T/|W|$
and {\it the eccentricity} $e= (1-(r_{\rm p}/r_{\rm eq})^2)^{1/2} $
where $r_{\rm p}$ and $r_{\rm eq}$ are the polar and equatorial
coordinate radius. We have computed the $T/|W|$ according to the
prescription of Friedman et al.  (1986):
\begin{eqnarray}
T/|W|&=&{{\Omega J/2}\over{\Omega J/2 + M_{\rm p} -M}}
\end{eqnarray}
where $\Omega$ is the angular velocity, $J$ the stellar angular
momentum, $M$ is the gravitational mass and $M_{\rm p}$ is the ``proper ''
internal energy.
A rotating configuration becomes unstable to a bar-mode formation if
the ratio of $T/|W|$ or the eccentricity exceed some critical value.
For incompressible Newtonian bodies, the critical values of $T/|W|$ and $e$
are 0.1375 and 0.8126 respectively. The Newtonian value of $T/|W|$ is
0.1275 (Bonazzola et al. 1996) for compressible bodies.

General relativity weakens the viscosity driven bar mode instability,
but the stabilizing effect is not very strong (Gondek-Rosi\'nska \&
Gourgoulhon 2002, Di Girolamo \& Vietri 2002, Bonazzola et al.
1996, 1998). The ratio $T/|W|$ for relativistic neutron star models is
$\sim 0.14$ and for highly relativistic ($GM/(R_{\rm circ}c^2)=0.25$)
uniformly rotating incompressible bodies is only $~30\%
$ higher than
the Newtonian value (Gondek-Rosi\'nska \& Gourgoulhon 2002). 
It was found by Gondek-Rosi\'nska \& Gourgoulhon
(2002) that for rigidly rotating homogeneous fluid bodies the
dependence of $(T/|W|)_{\rm crit}$ on the proper compactness parameter
$x := GM/(R_{\rm circ}c^2)$ can be very well approximated by the
function
$$(T/|W|)_{\rm crit}= (T/|W|)_{\rm crit,Newt}+0.148x(x+1).$$  

We locate the onset of instability (see section 4.2 for description of
our method) along several constant baryon mass sequences of uniformly
rotating axisymmetric strange quark stars described by different MIT
bag models (see section 3.) for the compaction parameter $GM/(R_{\rm
circ}c^2) = 0-0.25$.  To show the role of relativistic effects we plot
in Fig.~\ref{fig:TWMR} the ratio of $T/|W|$ as a function of the proper
compaction parameter at the secular instability points for strange
stars. All 3 models described in section 3 are represented by a solid
line.  We compare our results with those obtained by Gondek-Rosi\'nska
\& Gourgoulhon (2002) for relativistic incompressible bodies (dashed
line). We see the same dependence of the critical value of $(T/|W|)_{\rm
crit}$ on compactness in both cases.  In the Newtonian limit
($GM/(R_{\rm circ}c^2)=0$) the ratio of $(T/|W|)_{\rm crit}$ is 0.1375.

In Fig~\ref{fig:oblat} we plot the oblateness of the star (the ratio of
$r_{\rm p}/r_{\rm eq}$) versus the proper compaction parameter for
both MIT bag model strange stars (solid line) and incompressible fluid
bodies (dashed line). It was found by Gondek-Rosi\'nska \& Gourgoulhon
(2002) that the critical value of the eccentricity depends very weakly
on the degree of relativity and for $GM/(R_{\rm circ}{\rm c^2})=0.25$
is only $2\%
$ larger than the Newtonian value at the onset for the
secular bar mode instability. In the case of strange stars the
dependence is a little bit stronger. For massive strange stars the
density profile is no longer completely flat - for example the ratio
of central density to surface density for a strange star is $1.25$ and
$1.8$ for proper compaction parameters $0.1$ and $0.2$ respectively.

\subsection{Results for the standard MIT bag model}

Here we present results for the standard MIT bag model, SQS1, where the
mass of strange quark is taken to be $200~{\rm MeV/c^2}$, the coupling
constant 0.2 and the bag constant $B=56~{\rm MeV/fm^3}$.  We
 show that the results presented in this section can be applied to
the other MIT bag models of SS to a very good approximation (Sec 5.3).

\subsubsection{Constant baryon mass sequences}

\begin{figure*}
\includegraphics[width=0.9\columnwidth]{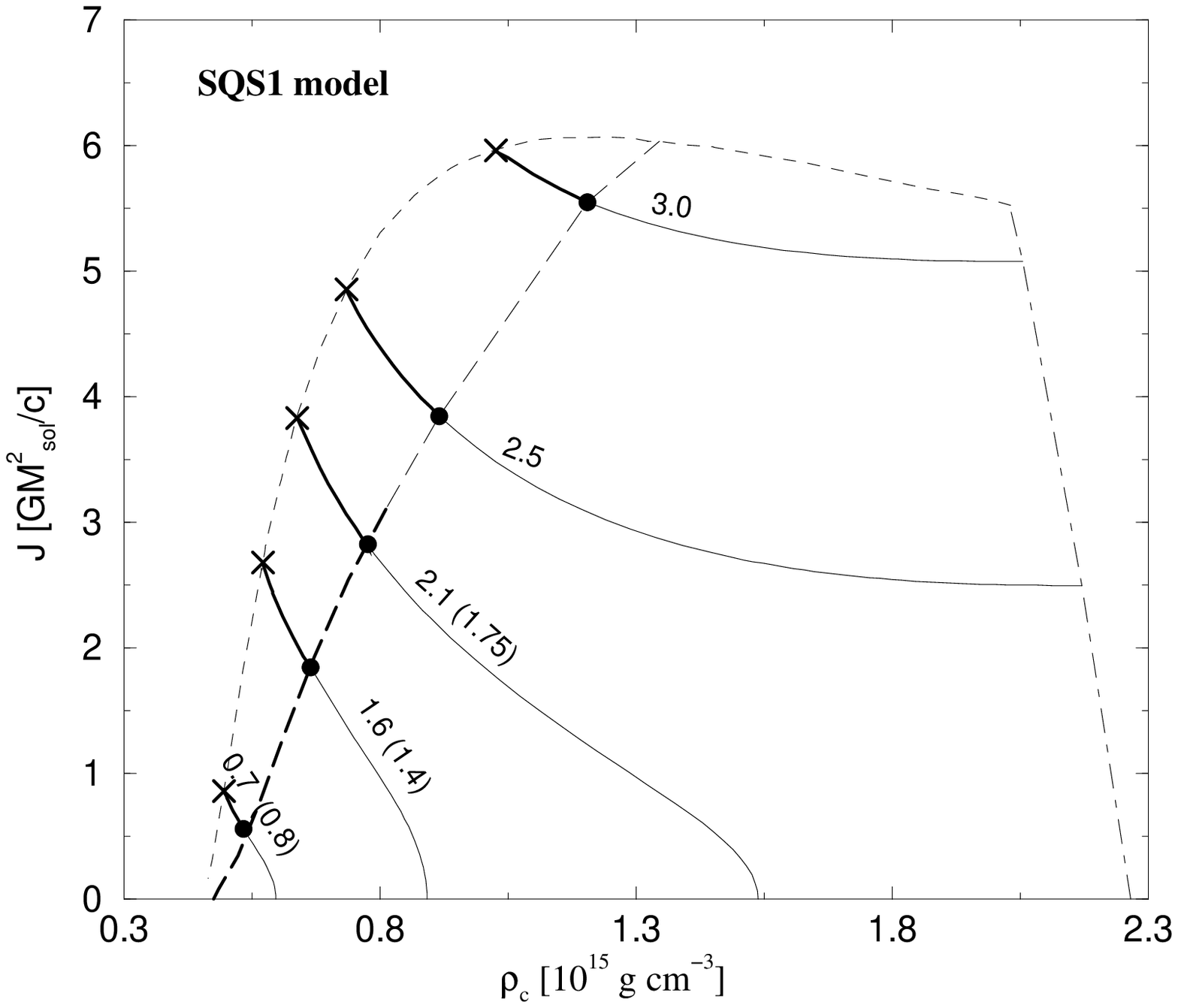} \qquad
\includegraphics[width=0.9\columnwidth]{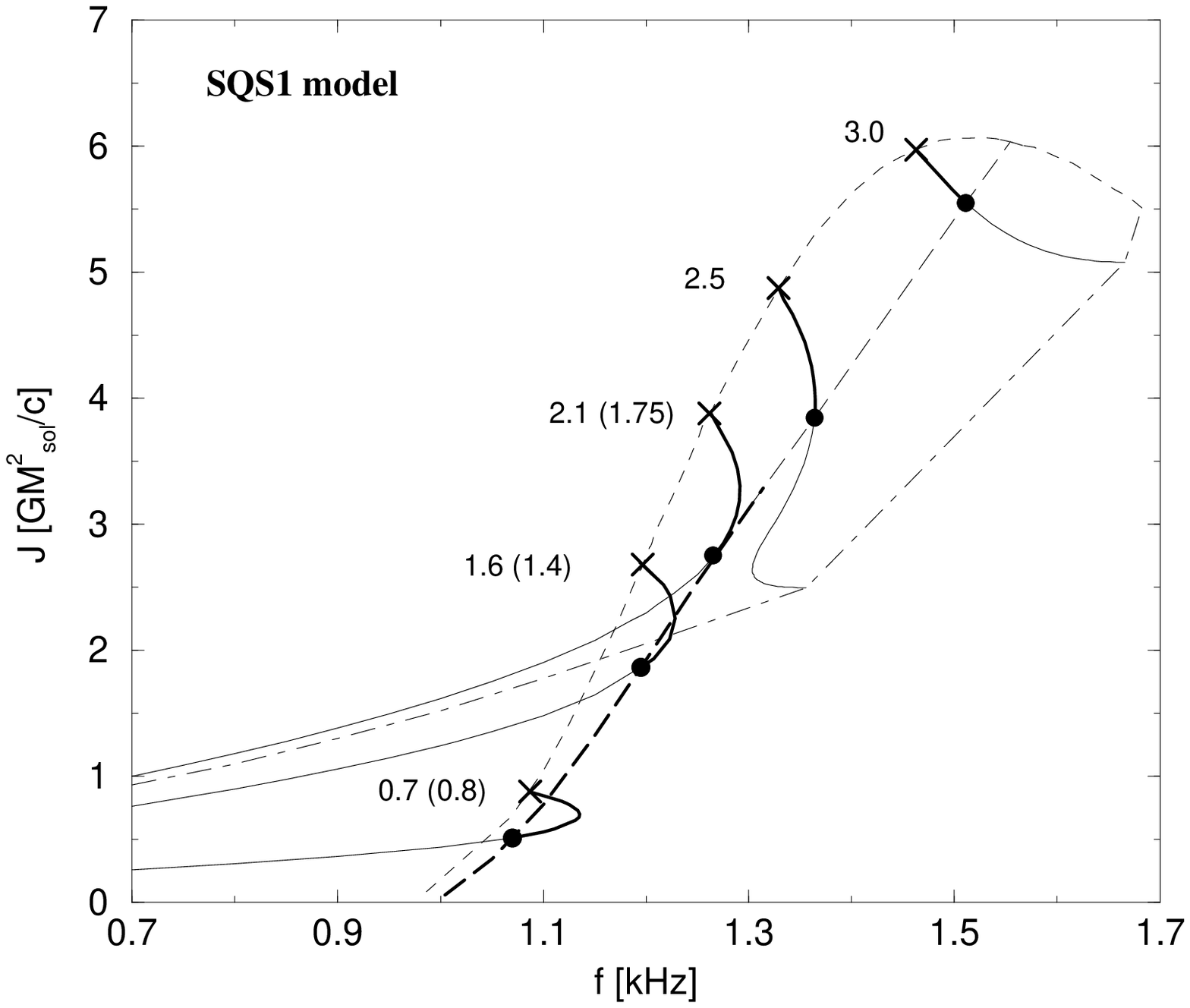}
\caption[]{Angular momentum as a function of central density (left)
and the rotation frequency $f=\Omega/(2\pi)$ (right) along sequences
of constant baryon mass (solid lines). Each solid line is labeled by
this baryon mass in solar mass units, as well as (in parentheses) by
the gravitational mass of the static configuration, if it exists.  The
dash-dotted line indicates stars marginally stable to axisymmetric
perturbations.  The Keplerian configurations are shown as thin
short-dashed line (also as crosses for each of considered
sequences). The long dashed line indicates stars marginally stable to
non-axisymmetric perturbations. Thick solid lines represent
the configurations secularly unstable to the bar mode formation taking into
account viscosity as a leading dissipative mechanism }
\label{fig:J}
\end{figure*}

We construct constant baryon mass sequences of rotating strange stars
described by the SQS1 model, i.e. the so-called evolutionary
sequences. This is relevant for the evolution of an isolated strange star
slowly losing its energy and angular momentum via gravitational and
electromagnetic radiation. In general we cannot neglect accretion on a
compact star (e.g. an accretion from supernova on a just born strange
star or from a companion in low mass X-ray binaries).

 We compute both {\it normal} (with $M_{\rm b}=0.8, 1.6, 2.1 M_\odot$) and
{\it supramassive} sequences (2.5, 3.0 $M_\odot$).  A sequence is called
normal if it terminates at the zero angular momentum limit with a
static, spherically symmetric solution, and it is called a
supramassive sequence if it does not.  The boundary between these two
sequences is the sequence with the maximum baryon mass of a static
configuration, which is $2.17 M_\odot$ for the SQS1 model.  We consider
configurations stable with respect to axisymmetric perturbations, i.e
for a supramassive constant baryon mass sequence a model is stable to
quasi-radial perturbations, if $\displaystyle \left( {\partial J \over
\partial \rho_{\rm c}} \right)_{M_{\rm b}} < 0$ (or $\displaystyle
\left( {\partial M \over \partial \rho_{\rm c}} \right)_{M_{\rm b}} <
0$) where $M_{\rm b}$ is the baryon mass of the star and $\rho_{\rm
c}$ is the central density of the star (Friedman, Ipser \& Sorkin
1988).

The dependence of angular momentum versus central density as well
versus rotational frequency for the constant baryon sequences is shown
in Fig.~\ref{fig:J}. Each sequence is represented by a solid line and is
labeled by the baryon mass as well as by the gravitational mass of the
static configuration (in parentheses), if it exists.  The
gravitational mass of a strange star rotating at the Keplerian limit
is only a few percent (up to $7\%
$) higher than $M$ of the static member of
the sequence (for the influence of rotation on the gravitational mass see
Figs. 2 in Gondek-Rosi\'nska et al. 2000b and Gondek-Rosi\'nska et al.
2001a).  Marginally stable configurations to quasi-radial perturbation
are shown as a dot-dashed line.  The angular momentum increases along
each sequence from $J=0$ for static configurations, or $J_ {\rm min}$
on the dash-dotted line for supramassive stars, to $J_{\rm max}$ for
the Keplerian configurations (thin short dashed line, as well as the
crosses for each considered sequence). The star in a sequence reaches
the mass-shedding limit if the velocity at the equator of a
rotating star is equal to the velocity of an orbiting particle at the
surface of the star.  It is worth noting (Fig.~\ref{fig:J}) that for
any given baryon mass, the maximal rotating configuration is not
mass-shedding one. The mass-shed configurations are reached due to the
increase of the equatorial radius relative to the deformation of the
rotating star.  Such a phenomenon was discussed by Zdunik et al.
(2000b) in the case of normal sequences of MIT bag model strange stars
and Gondek-Rosi\'nska et al. 2000 for the Dey et. al (1998) EOS. This
feature is characteristic of stars described by a self-bound linear
EOS. If a strange star is born with high angular momentum (close to
the maximum available) it will spin up, losing the angular momentum
(e.g by emitting of gravitational waves).

Marginally stable configurations to non-axisymmetric perturbation are
shown as a long-dashed line (as well as a filled circle for each
considered constant baryon mass sequence).  As mentioned in Sec.4, the
multi-domain spectral technique with surface-fitted coordinates that
we employ has some trouble when $r_{\rm p}/r_{\rm eq} < 1/3-0.5$
depending on the compaction parameter. To compute very oblate and
highly relativistic (with high compaction parameter) rotating
axisymmetric stationary configurations we use the Stergioulas \& Friedman
(1995) code, but we cannot locate the Jacobi-like bifurcation point in
this case.
The thin part of the long dashed line (for $f > 1.3~{\rm kHz}$ for the SQS1 model)
represents marginally stable configurations found in the following
way - we noticed that the dependence of central enthalpy $H_{\rm c}$ on
rotational frequency $f$($H_{\rm c}$ and $f$ are two input parameters
needed for finding a rotating star described by given equation of
state) at the onset of instability is very well approximated by the
following formula: $H_{\rm c}=0.47f_{[{\rm kHz}]}-0.45$ for SQS1 model.
We extrapolated this formula to higher rotational frequency in order
to find central enthalpy to compute the axisymmetric configuration.

Thick parts of the solid lines, between the filled circle and the cross,
correspond to configurations secularly unstable to the bar mode
formation.  These configurations can be good emitters of gravitational
waves.  However a strange quark star described by this model must
rotate very rapidly, with frequency higher than 1.1 kHz for $1.4 \
M_\odot$, to allow symmetry breaking.  If viscosity is high (see
discussion in Sec 6), an accreting strange star (belonging to a normal
sequence) in low mass X-ray binaries (LMXBs) will never reach the
mass-shedding limit. It can be accelerated to the critical frequency
(which can be higher than the Keplerian frequency - see Fig. 3),
corresponding to the marginally stable configuration to the bar mode
formation instead.

\subsubsection{Masses and rotational frequencies for secularly unstable strange quark stars}
\begin{figure}
\includegraphics[width=0.9\columnwidth]{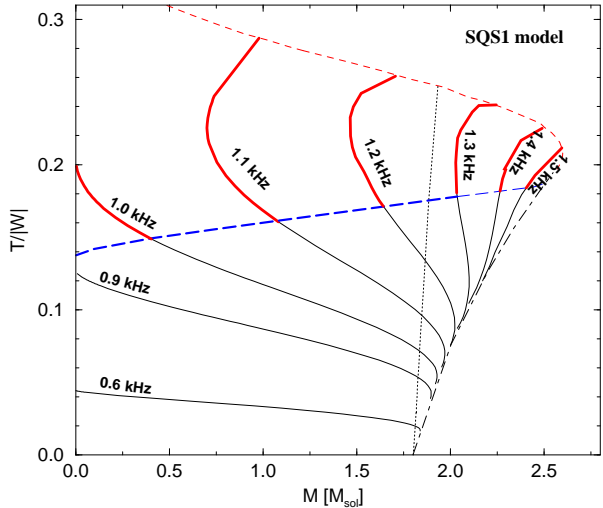}
\caption{ The ratio of the kinetic energy to the absolute value of the
gravitational potential energy as a function of gravitational mass for
the standard MIT bag model of strange quark stars. The Newtonian limit
is obtained for $M\to 0$ ($J \to 0$, $\rho_c\to\rho_0$) (a rotating
strange star can be considered as the Maclaurin spheroid). The solid
lines correspond to sequences of configurations rotating with constant
rotational frequency. The rotational frequency is labeled close to
each line.  On one end of each sequence there is a mass-shed
configuration (short-dashed line) and on the other end the last stable
configuration with respect to axisymmetric perturbations (dash-dotted
line). The dotted line separates normal strange stars from
supramassive strange stars. The meaning of the other lines is the same
as in Fig. 3. Configurations with $T/|W|$ higher than represented by
long-dashed line (e.g. thick parts of solid lines) are secularly
unstable to the bar mode formation taking into account viscosity as a
leading dissipative mechanism.}
\label{fig:TWM}
\end{figure} 

\begin{figure}
\includegraphics[width=0.9\columnwidth]{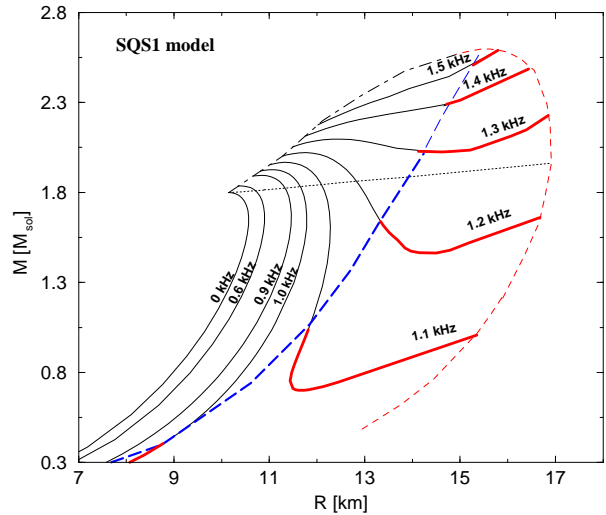}
\caption{Gravitational mass vs. radius for stars described by the
  standard MIT bag model. All lines have the same meaning as in
  Fig. ~\ref{fig:TWM}. Models located to the right of the long dashed
  line are secularly unstable to the viscosity driven bar mode formation.}
\label{fig:MR}
\end{figure}

The crucial question from the observational point of view is - for
which masses and frequencies the triaxial instability in the case of
strange stars can develop. In order to answer this question we plot
the value of $T/|W|$ versus gravitational mass for several constant
frequency $f=\Omega/2\pi$ sequences (thin solid lines) of uniformly
rotating strange stars in Figure~\ref{fig:TWM}.  On one end of each
sequence there is a mass-shedding configuration (with the lowest
central density in the sequence), and on the other end the last stable
configuration with respect to axisymmetric perturbations (the densest
object in the sequence). The dotted line separates normal strange
stars from supramassive strange stars.

We see that the ratio of $T/|W|$ for strange quark stars rotating at
the mass-shedding limit is very large and can be even higher than
$T/|W|=0.32$.  The $T/|W|$ is very high for all linear self-bound EOS
(see Gondek-Rosi\'nska et al., 2000b for different MIT bag models and
Gondek-Rosi\'nska et al., 2000a for strange stars described by the Dey
model). The values of $T/|W|$ for strange stars are much larger than
for any neutron star models (for which the maximum $T/|W|$ ranges
typically from 0.1 to 0.14 - see e.g.  Cook et al., 1994). In contrast
to neutron stars, the $T/|W|$ for strange stars increases with
decreasing stellar mass for all models.  Very high values of $T/|W|$
are characteristic not only of supramassive stars but also of
rapidly rotating strange stars with small and moderate masses. The
large value of $T/|W|$ results from a flat density profile combined
with strong equatorial flattening of rapidly rotating strange stars.
The long-dashed line represents the onset of the secular bar mode
instability (critical values of $T/|W|$ and $e$). Configurations with
$T/|W|$ higher than represented by this line are unstable with respect
to non-axisymmetric perturbation and can be good emitters of
gravitational waves (thick solid lines).

From Figs~\ref{fig:TWM} and ~\ref{fig:MR} we see that strange stars could
break their symmetry far away from the Keplerian limit and for broad
ranges of masses and rotational frequencies. The critical value of
$T/|W|$ is in the range $0.1375-0.18$ for very low mass rotating
strange stars (in the Newtonian limit very well approximated by
Maclaurin spheroids) and supramassive stars respectively.  Those
values are much lower than the Newtonian value of the dynamical
instability against bar-mode deformation, 0.27 (according to our
knowledge no stability analysis has been performed yet for uniformly
rotating stars in general relativity). Existence of a crust only
marginally affects our results. It has little influence on location of
the viscosity driven instability points. According to Zdunik, Haensel
and Gourgoulhon (2001) the maximal values of $T/|W|$ at mass-shedding
limit are only $20 \%
$ lower for rapidly rotating strange stars with a
maximal crust than the corresponding values for bare strange stars
described by the same MIT bag model.  From Fig. 3 we see that a high
rotational frequency, higher than $0.95~{\rm kHz}$ ($>1.1~{\rm kHz}$
  for 1.4~$M_\odot$), is needed for a star described by the standard
  MIT bag model to reach the instability region. If we consider a model
  with $m_{\rm s}=200~{\rm MeV/c^2}$, but with high coupling
  constants $\alpha=0.6$ and low bag constants $B=36.4~{\rm
    MeV/fm^3}$ then the critical frequency for which the viscosity driven
  bar mode instability could occur is always $>0.77~{\rm kHz}$ and is
  $0.88~{\rm kHz}$ for a canonical strange star.

\subsection{Results for other MIT bag models}

\begin{figure}
\includegraphics[width=0.9\columnwidth]{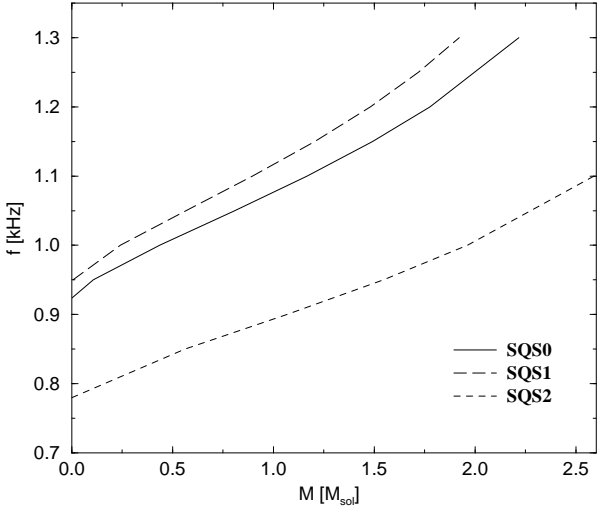}
\caption{Rotational frequency vs. gravitational mass for secular (
 viscosity driven) instability configurations described by different
 MIT bag models. Stars rotating with higher frequency than critical
 one (shown by each curve) are secularly unstable to non-axisymmetric
 perturbations.}
\label{fig:fM}
\end{figure}

\begin{figure}
\includegraphics[width=0.9\columnwidth]{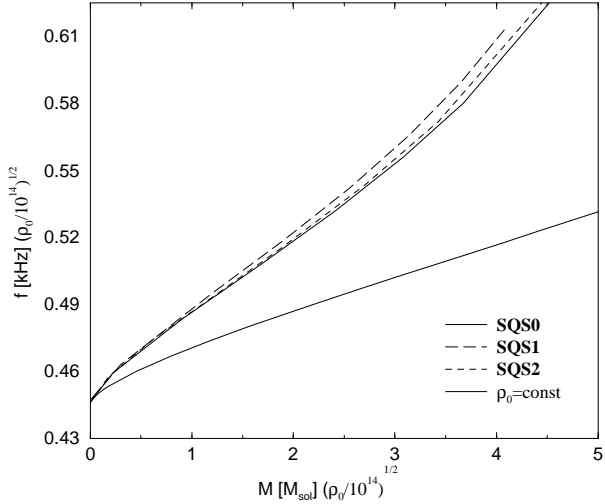}
\caption{Rescaled Figure ~\ref{fig:fM} to see how results depend on
the $a$ parameter at equation of state given by eq. (1). The onset
points of stability for incompressible fluid bodies is shown for
comparison (thin solid line). All other lines and symbols have the
same meaning as in Fig. ~\ref{fig:fM}.}
\label{fig:fMs}
\end{figure}

In Fig ~\ref{fig:fM} we show gravitational masses and rotational frequency
for which triaxial instabilities in rotating strange quark stars can
develop. Each line corresponds to the onset of instability for
particular MIT bag model described in Sec 3.  Strange stars rotating
with a frequency higher than represented by this line are unstable with
respect to non-axisymmetric perturbations and can be good emitters of
gravitational waves. We see that viscosity driven instability develops
for a broad range of gravitational masses of strange stars (from that
of a planetoid to about three solar masses) independently of the model.
However, a high rotational frequency (always higher than the frequency
$\sim 0.64~{\rm kHz}$ of the fastest observed pulsar) is needed for a
star to reach the instability region. The higher the gravitational mass 
the higher is the frequency needed to form a ``bar-shaped star''. The most
comfortable situation is for MIT bag models with low MIT bag
constant (e.g. SQS2) and for small stellar masses. However one should 
keep in mind that detailed simulations of supernovae
do not predict remnant masses less than $ ~1.2 M_\odot$ (Timmes et al. 1996).

It was shown that strange stars described by different models of
strange matter can be very well approximated (with accuracy $1-2\%
$)
by a linear function (Zdunik 2000, Gondek-Rosi\'nska et al. 2000a)
$$P=a(\rho-\rho_0)c^2, \eqno (1)$$ where $P$ is the pressure, $\rho$
the mass-energy density, and $c$ the speed of light. Both $\rho_0$ and
$a$ are functions of the physical constants $B$, $m_{\rm s}$ and
$\alpha$ for the MIT bag model.  In general, eq. (1) corresponds to
self-bound matter with density $\rho_0$ at zero pressure and with a
fixed sound velocity ($\sqrt{a}\,c$) at all pressures.  For the MIT
bag model the density of strange matter at zero pressure is in the
range $\sim 3 \times 10^{14}-6.4 \times 10^{14}\ {\rm g\ cm^{-3}}$ and
$a$ between 0.289 (for a strange quark mass of $m_{\rm s} c^2$= 250
MeV) and 1/3 (for massless quarks).  For a fixed value of $a$, all
stellar parameters are subject to scaling relations with appropriate
powers of $\rho_0$, $f \propto \rho_0^{1/2}$, where $f$ denotes either
of the frequencies (rotational or orbital), and $M, R \propto
\rho_0^{-1/2}$ (see e.g Witten 1984; Zdunik 2000). The scaling law for
stellar parameters with $a$ depends on rotational frequency
(Gondek-Rosi\'nska et al 2003), for approximate scalings with $a$ at
the limiting case of rotation at mass-shedding frequencies see
Stergioulas, Klu\'zniak \& Bulik (1999).
 
In Fig.~\ref{fig:fMs} the dependence of rotational frequency vs. mass
for marginally stable strange stars to non-axisymmetric perturbations
(Fig. ~\ref{fig:fM}) has been scaled out with appropriate powers of
$\rho_0=10^{14}\ {\rm g\ cm^{-3}}$.  We see that the function $f(M)$
for onset points of instability very weakly depends on $a$ ($a$
changes not very much in MIT bag model).  The results obtained for one
model can be used for other MIT bag models with the use of the scaling
with $\rho_0$. The critical points for incompressible fluid stars with
$\rho_0={\rm const}=10^{14}\ [{\rm g\ cm^{-3}}]$ are shown for comparison
(thin solid line).

From Fig~\ref{fig:fMs} we see that independently of any details of SS
EOS, the lower the stellar mass  the lower rotational frequency $f_{\rm
crit}$ needed to reach the instability region. Taking this into account
we can find the absolute lower limit on rotational frequency of SS
when viscosity driven instability can develop.  In the Newtonian
limit, when $M\to 0$, $J \to 0$, $\rho_c\to\rho_0$, a rotating strange
star can be considered as the Maclaurin spheroid for which the
rotational frequency depends only on density and eccentricity (or
$T/|W|$) (Chandrasekhar 1969). The square of the angular velocity at
the onset point of viscosity driven instability ($e=0.8126$) is
$$ {{\Omega_{\rm crit}}^2 \over {\pi G \rho_0}}=0.3742, $$ in other
words for a strange star with gravitational mass $M\to 0$ the
frequency ${f_{\rm crit} \over \sqrt{\rho_{0,14}}} \to 0.4458\ [{\rm
kHz}]$.  The lowest value of $f_{\rm crit}$ is obtained for the lowest
value of $\rho_0$ (fulfilling the condition of stability neutrons to
spontaneous fusion into droplets of u and d quarks as well as that
energy per baryon for strange quark matter must be below 930.4
MeV). For the MIT bag model we can find that the lowest rotational
frequency at which triaxial instability can develop is $\sim
0.75$~kHz.

\begin{figure}
\includegraphics[width=0.9\columnwidth]{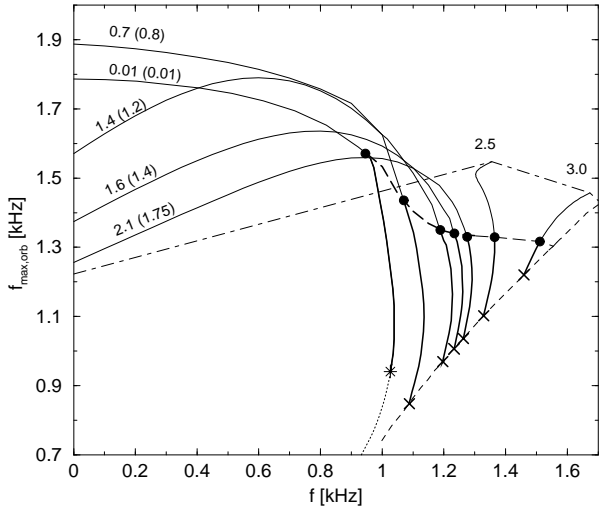}
\caption[]{The secularly unstable configurations to the bar mode formation
  (thick solid lines between a heavy black circle and a cross) on the
  maximum orbital frequency versus frequency of the rotation plane. The
  long dashed line indicates stars marginally stable to
  nonaxisymmetric perturbations taking into account viscosity as a
  leading dissipative mechanism. One sequence for a very low mass star
  (with $M_{\rm b}=0.01 M_\odot$) is shown, the critical point on this
  sequence for Newtonian dynamical instability to non-axisymmetric
  perturbations is indicated by an asterisk and dynamically unstable
  configurations are denoted with the dotted line.  }
\label{fig:fmsfb}
\end{figure}

\subsubsection{Lower limits on the maximum orbital frequency around
rotating strange quark stars and kilohertz QPOs}

In this section we show that triaxial instability is important for
interpretation of maximal orbital frequency of rapidly rotating
strange stars in LMXBs. We focus on the MIT bag model but our results are
robust for other linear self-bound EOS.  The spin-up of a relativistic
star crucially depends on the maximum orbital frequency around it, as
do a host of other high energy accretion phenomena in low mass X-ray
binaries, including quasi-periodic oscillations (QPOs) in the X-ray
flux.  Klu{\'z}niak et al.(1990) suggested that the mass of a neutron
star may be derived by observing the maximum orbital frequency, if it
occurs in the marginally stable orbit, and that the low frequency QPOs
occurring in X-ray pulsars will have their counterpart in LMXBs, at
frequencies in the kHz range. Such kHz QPOs have indeed been
discovered and their frequency used to derive mass values of about
$2M_\odot$, under the stated assumption (Kaaret et al. 1997, Zhang
et al. 1998, Klu\'zniak 1998, Bulik 2000). In fact, the QPO
frequency may correspond to a larger orbit, and hence a smaller mass
(Klu\'zniak \& Abramowicz 2002). The question of whether quark stars may
have maximum orbital frequencies as low as 1.07 kHz in 4U 1820-30
(reported by Zhang et al. 1998) has also been investigated (Bulik et
al.  1999a,b, Stergioulas et al. 1999, Zdunik et al. 2000a,b,
Gondek-Rosi\'nska et al. 2001a,b).  The main result of these
considerations was that the lowest orbital frequency at the innermost
marginally stable orbit (ISCO) is attained either for non-rotating
massive configurations close to their maximum mass limit, or for
configurations at the Keplerian mass-shedding limit. In order to
obtain an orbital frequency in the ISCO as low as 1.07 kHz for slowly
rotating strange stars one has to consider a specific set of (rather
extreme) MIT bag model parameters (Zdunik et al. 2000a, b) allowing
gravitational masses higher than 2.2 $M_\odot$.  In contrast, for
stars rotating close to the Keplerian limit, a low orbital frequency
at the ISCO can be obtained for a broad range of stellar masses
(Stergioulas et al., 1999, see Gondek-Rosi\'nska et al. 2001a for
discussion).

We present the dependence of the maximum orbital frequency (the
frequency at the marginally stable orbit if it exists or the Keplerian
frequency at the stellar surface, $f_{\rm orb,max}= f_{\rm orb}(R_*)$)
on the frequency of rotation f for the set of constant baryon mass
sequences of strange stars in Fig.~\ref{fig:fmsfb}. Marginally stable
orbits exist for a broad range of masses and rotational rates (see
e.g. Fig 1 in Gondek-Rosi\'nska \& Klu\'zniak 2002). As it turns out,
the very low mass limit, represented in Figure~\ref{fig:fmsfb} by the
case of $M_{\rm b}=0.01 M_\odot$, is described well by the Maclaurin
spheroid for which analytic formulae for orbital frequencies have been
derived and found to agree with the relativistic numerical models
(Amsterdamski et al., 2002).  For this sequence, an ISCO exists only
for rotation rates larger than $\sim 950 \ {\rm Hz} $. We indicate the
onset of the (Newtonian) dynamical instability to non-axisymmetric
perturbations by an asterisk. In the Newtonian limit, the gap between
the marginally stable orbit and the stellar surface is produced by the
oblateness of the rapidly rotating low-mass quark star (Klu\'zniak,
Bulik \& Gondek-Rosi\'nska, 2001; Amsterdamski et al., 2002; Zdunik \&
Gourgoulhon 2001). We see that very rapidly rotating strange stars are
secularly unstable to bar mode formation if the viscosity is high enough.
The lowest frequency in the marginally stable orbit is
obtained for slowly rotating massive strange stars or a marginally
stable configuration to nonaxisymmetric perturbation (long dashed
line). If the mass of the strange quark is relatively small, $m_{\rm
s}c^2 \sim 100\ {\rm MeV}$, then the r-modes instability could play an
important role as an emission mechanism of gravitational waves and a
mechanism spinning down rotation rates of strange stars (Andersson
et. al. 2002).

\section{Rotating strange quark stars as sources of gravitational
waves}

The interesting question is whether the viscosity-driven instability
domain can be reached by any astrophysical process.  In section 5 we
located the onset of instability (when the frequency of an $l=-m$ mode
goes through zero in the corotating frame) along constant baryon
sequences of rotating strange quark stars.  This necessary condition
yields the critical angular velocity for a given gravitational mass of a
star. From our numerical results, the rotation frequency has to be
above $0.8-1.2 {\rm\ kHz}$ for strange quark stars described by the MIT
bag model of normal quark matter (any kind of superconductivity being
excluded). The instability can develop for a broad range of
gravitational masses of rotating SS.

The above results were obtained under the assumption of infinite shear
viscosity.  The necessary requirement is that the 'real' viscosity has
to be high enough to dissipate energy and let the instability develop
along the Jacobi path.  

As we mentioned in Sec 2 if both viscosity and gravitational radiation
are operative, they act against each other.  Since the
viscosity-driven mode is triaxial it tends to be damped by
gravitational-wave emission, and since the gravitational-wave driven
mode involves differential rotation, it is damped by viscosity.  The
instability can be important if the typical instability growth rate
($\tau_\eta$ shear viscosity timescale) is faster than any damping
mechanisms. 

It is important to emphasize that the spin evolution of strange stars
depends on many different physical mechanism, concerning
e.g. viscosities of dense matter, cooling evolution, the influence of
magnetic field on instabilities etc, for which our understanding is still
unsatisfactory.  Below we consider different timescales (shear
viscosity, bulk viscosity, cooling timescale, gravitational radiation)
and provide a crude estimate of the stellar temperature for which
viscous effects could dominate. Viscosity driven instability is
divergence free so bulk viscosity does not influence it. The
shear and bulk viscosity (axial modes are divergence free but mode
coupling should be taken into account) of strange quark star matter is
able to suppress the growth of the CFS-instability except when the
star passes through a certain temperature window.  The $l=m=2$
r-mode is expected to lead to the strongest CFS-instability so we
will take into account only this mode in our discussion.

\subsection{Viscous timescales connected to shear viscosity}
\label{subsect:shear.timescales}

Shear viscosity of normal SQM has been calculated by Heiselberg and Pethick
(1993):
\begin{equation}
\eta=5\times 10^{15}\;\left({0.1\over \alpha}\right)
\left({\rho\over \rho_0}\right)^{14/9}\;
{\rm T_9}^{-5/3}~{\rm g~cm^{-1}~s^{-1}}~,
\label{eta.sqm.formula}
\end{equation}
where ${\rm T_9} = {\rm T/10^9}~$K.
Assuming $\rho=4\rho_0$, $\alpha_{\rm s}=0.1$, and $R=10~{\rm km}$,
we estimate  the viscous timescale for a SS as
\begin{equation}
\tau_\eta={\rho R^2\over \eta}= 5.3\times 10^8~{\rm T_9}^{5/3}~{\rm s}~.
\label{tau.eta.SS}
\end{equation}
This is characteristic timescale of the angular momentum transport within
a strange star.

\subsection{Viscous timescales connected with bulk viscosity}
\label{subsect:bulk.timescales}
Bulk viscosity $\zeta$ of the SQM is due to the non-equilibrium 
non-leptonic strangeness changing process
\begin{equation}
u + d \rightleftharpoons s +u~.
\label{process.ud.su}
\end{equation}
Local compression or decompression of the SQM drives it from the flavor 
equilibrium, because chemical potentials for the ultrarelativistic up and 
down quarks have weaker density dependence than the chemical potential of 
the massive strange quarks. Therefore, a baryon density perturbation implies 
$\mu_s\neq \mu_d$.  The non-equilibrium reactions, changing the concentrations 
of the down and strange quarks, and driving the system toward  equilibrium, 
are irreversible, and therefore imply dissipation. The dissipation rate 
depends sensitively on the strange quark mass. Of course, dissipation vanishes 
in the (unphysical) limit of $m_s\longrightarrow 0$, where the flavor 
symmetry is enforced. The dissipation rate in a  pulsating  SQM depends on 
the temperature, density, frequency of pulsations $\omega$, and on the density 
oscillation amplitude $\delta n_{\rm b}$. The time-averaged dissipation rate can 
be represented by the (effective) bulk viscosity of SQM. The formula for $\zeta$ 
was derived by Madsen (1992), and can be written as 
\begin{eqnarray}
\zeta &=&  9\times10^{27}~\left({m_s c^2 \over 100~{\rm MeV}}\right)^4
\left( {\omega\over 10^4~{\rm s^{-1}}}\right)^{-2}\; 
{\rho\over \rho_0} {\rm T}_9^2\cr
&~&\times\left[1  + b {\rm T_9}^4(\omega / 10^4~{\rm s}^{-1})^{2} \right]^{-1}~,
\label{eq:zeta.general}
\end{eqnarray}
where the dimensionless coefficient $b$  is proportional 
to  the coefficient $\beta$ of Madsen (1992), 
and the terms depending on $\delta n_{\rm b}$ were assumed to be 
unimportant (see also Andersson, Jones, Kokkotas, 2002).
The second term in square brackets is important for ${\rm T}\ga 10^{10}~$K, 
where it leads to the quenching  of $\zeta$, opening in this way the possibility of 
the $ r$-mode instability in very hot rotating strange stars 
(Madsen 2000).
 
The terms  depending on $\delta n_{\rm b}$  become 
important  when 
\begin{equation}
{\delta n_{\rm b}\over n_{\rm b}}{m_s c^2 \over \mu_d T_9}>2~.
\label{eq:cond.dens.ampl}
\end{equation}
Such a condition can be realized at low temperatures  ${\rm T}\la 10^8$ K only 
provided $\delta n_{\rm b}/n_{\rm b}$ in the r-mode is sizable, 
$\delta n_{\rm b}/ n_{\rm b}\ga  0.1$: this could happen only for a 
large-amplitude r-mode instability. In what follows, the second term 
in the square brackets will be omitted.

Consider now  ${\rm T}<5\times 10^{9}~$K, when the 
square bracket in Eq.\ (\ref{eq:zeta.general}) can be replaced by one. 
Using results of Madsen (2000) and assuming a mean density of 
strange star $3\rho_0$, we get the following 
rough estimate for the viscous 
damping timescale of the r-mode, 
\begin{eqnarray}
{\rm T}&<& 5\times 10^9~{\rm K}~:\cr\cr 
&~&\tau_\zeta \simeq  
4\;\left({m_s c^2 \over 100~{\rm MeV}}\right)^{-4} 
\;\left({f\over 1000~{\rm Hz}}\right)^{-2}\;
{\rm T}_9^{-2}
~{\rm s}~.
\label{eq:tau.zeta}
\end{eqnarray}
The damping timescale depends strongly on  $m_s$, which is a {\it renormalized} 
strange quark mass (see Sect. 3.1).  

\subsection{Cooling timescale}
 
The SS cools rapidly, with
 \begin{equation}
\tau_{\rm cool}\equiv {{\rm T}\over \dot{\rm T}}=
{c_V {\rm T}\over Q_\nu}~,
\label{tau.cool}
\end{equation}
where $c_V$ is the specific heat of SQM (at a fixed volume),
and $Q_\nu$ is the neutrino emissivity. Eq.\ (\ref{tau.cool}) can be
easily integrated to give the time (age) needed to cool SS to  ${\rm T_8
\equiv T/10^8~}$K,
\begin{equation}
t_{\rm cool}={10^8\over {\rm T_8^4}}~{\rm s}~.
\label{t.cool.T8}
\end{equation}

\subsection{Damping of the Jacobi-like instability}

Lindblom \& Detweiler (1977) have shown that if both
viscosity and gravitational radiation reactions are taken
into account, rotating stars can be stable beyond the 
Jacobi bifurcation point located in Sec.~\ref{s:results}. 
This stability can be seen as resulting from a  
competition between the CFS Dedekind-like mode and the 
Jacobi-like mode. It is governed by the ratio (deduced from
Eq.~(8) of Lindblom \& Detweiler 1977)
\begin{equation} \label{e:X_incomp}
      X=4.41 \times 10^{-13} 
      \left({\nu\over {\rm cm}^2 {\rm s}^{-1}} \right)
      \left({ R \over 10{\rm\ km}} \right) ^2
      \left( { M \over 1.4\, M_\odot}\right) ^3,
\end{equation}      
  where $\nu=\eta/\rho$ is the kinematic shear viscosity.
If $X\gg 1$ the star is unstable with respect to the
Jacobi-like mode (viscosity-driven instability), 
whereas if $X\ll 1$, it is unstable with
respect to the Dedekind-like mode (CFS instability). 
In the intermediate regime, $X\sim 1$, the star
is stable well beyond the bifurcation point. 
The formula (\ref{e:X_incomp}) has been established
for incompressible stars. Taking into account the compressibility
does not change it significantly, since for polytropes
of adiabatic index $\gamma$ this merely
adds a factor $(\gamma-6/5)^2/(\gamma-1)^2$ in front of
it (Lai \& Shapiro 1995).

 Putting in Eq.~(\ref{e:X_incomp}) 
the value of $\nu$ deduced from  the shear
viscosity given by Eq.~(\ref{eta.sqm.formula}) leads
to
\begin{eqnarray}
      X & =& 2 \times 10^{-8} 
      \left({0.1\over \alpha } \right)
      \left({\rho\over \rho_0}\right)^{5/9}
      \left({10^9 {\rm \ K}\over T}\right)^{5/3} \nonumber \\
      & & \times \left({ R \over 10{\rm\ km}} \right) ^2
      \left( { M \over 1.4\, M_\odot}\right) ^3.
\end{eqnarray}      
The fact that $X\ll 1$ for astrophysicaly interesting temperatures T$>
10^5$K (see eq. \ref{t.cool.T8}) means that the viscosity-driven
instability cannot develop in rotating strange stars described by the MIT
bag model, even if a star rotates rapidly with $T/|W| > (T/|W|)_{\rm crit}$.

\subsection{The instability windows for r-modes}

The CFS-instability of r-modes in strange stars has been studied by
Madsen (2000) and Andersson et al. (2002) in the Newtonian limit. They
found the critical rotation frequency for a given stellar model as a
function of temperature solving the following equation
\begin{equation}
{1\over \tau_r}+{1\over \tau_{\eta}}+{1\over \tau_{\zeta}}=0,
\label{t.r}
\end{equation}
where $\tau_r\sim -22 (47)\left({M \over 1.4
M_\odot}\right)^{-1}\left({R\over{\rm 10
km}}\right)^{-4}\left({f\over{\rm kHz}}\right)^{-6} {\rm s}$ (Kokkotas
\& Stergioulas 1999) is the r-mode  growth  time-scale, taking into
account only the leading current multipole and assuming a constant
density profile (or n=1 polytrop). According to this study, the
instability operates in a very different temperature window than for
NS (see Fig 1 in Madsen 2000 for the dependence of critical spin frequency
on internal temperature for the MIT bag model strange star with $m_{\rm
s}c^2=100$ and 200 MeV). Since for SS bulk viscosity becomes weak at
both very low and very high temperatures, the r-mode instability
could occur in temperature windows ${\rm T} > 10^{10}$~K ($4 \
10^9$~K) and ${\rm T} < \ 10^7$~K ($2\ 10^8$~K) for $m_{\rm
s}c^2=200$~MeV, shown as the hatched area in our Fig.~9 (for $m_{\rm
s}c^2=100$~MeV above dotted line), if the cooling time is long enough.
Andersson et al. (2002) suggested that the r-modes in a strange star
never grow to large amplitudes and emit a persistent gravitational
wave signal.

\subsection{Astrophysical scenarios}

Let us consider two evolutionary scenarios for rotating strange stars.
The first scenario refers to a newly born, rapidly rotating SS,
cooling due to neutrino emission. The second scenario is that of a SS in a
close binary system, which heats and spins-up due to accretion of
matter. In both cases we assume that strange quark matter (SQM) is
normal (non-superconducting). 

There are important differences between the two astrophysical
scenarios mentioned above.  The first one is that the energy radiated
by a newborn relativistic star is supplied by its rotational kinetic
energy, if it acquired a sufficiently large amount of rotational
energy during its formation. For an accreting compact star it is
supplied by the accreted matter. An old strange star in a close binary
system, accreting matter from its companion, may be spun up until the
ratio of $T/|W|$ is high enough to allow for the symmetry breaking. A
steady state regime is achieved when the total accreted angular
momentum is evacuated via gravitational radiation (Wagoner 1984,
Papaloizou \& Pringle 1978), this process being known as {\it forced
gravitational emission} or the {\it Wagoner mechanism}.
In this case the frequency of the emitted gravitational waves depends upon
the type of triaxial instability.
For the main $l=m=2$ r-mode the frequency of the emerging gravitational
waves is (see e.g. Stergioulas 1998)
$$f_{\rm gw}^{\rm r-mode} = {4\over 3}f.$$
For the Jacobi-like bar mode instability, the frequency of
gravitational waves is simply twice the rotational frequency:
$$f_{\rm gw}^{\rm visc} = 2 f.$$
If, in addition to the measure of $f_{\rm gw}$ by a gravitational
wave detector, one knows $f$ from electromagnetic observations
(e.g. X-ray pulsations), one can determine the kind of secular rotational
instabilities acting on the star, and consequently have some insight
into the value of the physical parameters (viscosity, temperature) inside
the compact object.

\subsubsection{Newly born rapidly rotating strange star}
\label{subsect:newborn.SS}
A SS is born from a proto neutron star with an initial temperature of
$\sim 10^{11}~$K, and cools via neutrino emission (shown in a schematic
way as the N track in Fig. 9). Of two basic types of SQM viscosity, the
bulk one is crucial for stabilizing rapid rotation with respect to the
r-mode instability (which develops on a timescale $\tau_r$).  The
region of the r-mode instability in the $f-{\rm T}$ plane strongly
depends on the value of $m_{\rm s}$, because $\tau_\zeta\propto m_{\rm
s}^{-4}$. Had we taken $m_{\rm s} c^2 =100~$MeV, the r-mode
instability of newly born SS would extend to much lower T and $f$.  We
use renormalized s quark mass $m_s c^2=200~$MeV 
consistent with particle data tables (see Sect. 3.1). Then the
r-mode is unstable within the right hatched strip in
Fig. ~\ref{fig:fT}. Due to the strong temperature dependence of
$\tau_\zeta$, stability of rotation is regained as soon as ${\rm T}$
decreases below $10^{10}$~K, which takes only a few seconds (see
eq. \ref{t.cool.T8}). A rapidly cooling young hot strange star would
not be spun down significantly passing the right r-mode instability
window.

At ${\rm T}=10^8~$K, $\tau_\eta \sim 10^7~ {\rm s}\ll t_{\rm
cool}(10^8~{\rm K})= 10^8$~s and a star could rotate rigidly.  Since
$X<<1$ (due to a very short gravitational damping time-scale) the viscosity
driven instability cannot develop even if the newly born SS left the
r-mode instability strip with $T/|W| >(T/|W|)_{\rm crit}$.  

Several years after its birth a star will reach the low-T (left) (when
temperature is a few times $ 10^7- 10^8$~K) r-mode instability
strip. During its cooling it will lose the excess of angular momentum
via GW radiation until it leaves the low-T r-mode instability window.

\subsubsection{Spin evolution of an old accreting strange star}
\label{subsect:acc.triax}
\begin{figure}
\includegraphics[width=0.9\columnwidth]{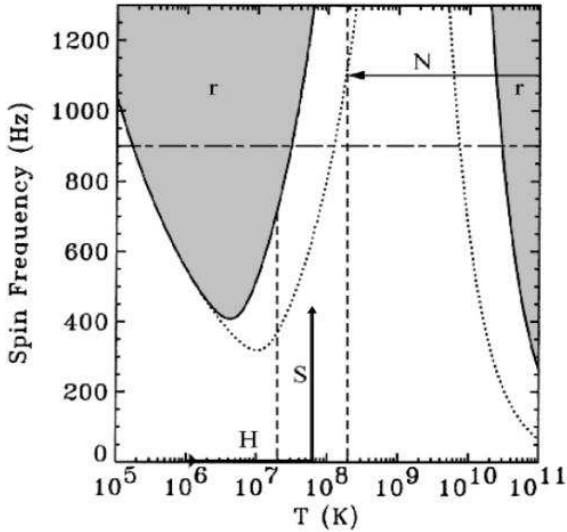}
\caption[]{ Regions of the r-mode instability (hatched region with
r-label) for a canonical mass strange star assuming $m_s c^2=200~$MeV
in the rotational frequency-temperature plane.  The r-mode instability
domain for $m_{\rm s} c^2=100~$MeV is shown for comparison (regions
above dotted lines). The r-mode instability windows are taken from
Madsen 2000.  The short-long dashed horizontal line corresponds to the
critical rotation rate for viscosity driven instability assuming
canonical stellar mass, $m_s c^2=200~$MeV and low MIT bag constant
$40$MeV/fm$^3$. The H-S tract corresponds to the evolution of an old
accreting SS and N tract to a newly born strange star.}
\label{fig:fT}
\end{figure}

Consider an old SS which starts to accrete matter in a close binary
system. Its initial internal temperature was very low, but energy
release associated with accretion will induce heating of SS interior.
To be specific, we consider accretion of a hydrogen-rich plasma.
There are two main channels of heating by accretion. The first one is connected
to plasma falling onto a bar SS surface, and the second one involves
formation of the crust and thermonuclear burning of compressed plasma
(Haensel \& Zdunik 1991). In both cases, additional heat release occurs
at the SQM core edge, where absorption and deconfinement  of protons
is associated with exothermic irreversible reactions $p\longrightarrow
2u + d$, $u+d\longrightarrow s +u$. The energy release per one deconfined
proton is $q_{\rm s}$. In what follows we will assume
$q_{\rm s}=50~$MeV. After some $10^{5}$~yrs SS,  accreting matter at a constant
rate $\dot{M}$, it will reach the steady thermal state, in which heating due to
accretion and deconfinement is balanced with neutrino losses from the SS core
(see Fujimoto et al. 1984). Notice that at a fixed  accretion rate (which will
be expressed in the units of $10^{-8}~{\rm M_\odot~yr^{-1}}$), $\dot{M}_{-8}$,
reaching of the steady state corresponds to accretion of only
 $10^{-3}\;\dot{M}_{-8}\;M_\odot$ (H-track in Fig.  ~\ref{fig:fT}).  
Equating heat gains and losses, one can estimate internal temperature
of accreting SS (Haensel \& Zdunik 1991),
\begin{equation}
{\rm T}=1.1\times 10^8\;\left(\xi M_*
\dot{M}_{-8}\right)^{1/6}\;R_6^{-2/3}~{\rm K}~,
\label{T.acc.SS}
\end{equation}
where $M_*=M/M_\odot$ and $R_6=R/10^6~{\rm cm}$. The
parameter $\xi$ varies from $\xi=0.2$ in the limiting case when the heating
results exclusively from the deconfinement process, to $\xi=1.2$
when all gravitational energy release by infalling protons flows
into the SS core (Haensel \& Zdunik 1991). Putting $M_*=1.6$ and
$R_6=1$, $\xi=0.2$, we get ${\rm T}=0.9\times 10^8$ K.

Further accretion of some $(0.3-0.4)\;M_\odot$ during some $5\times
10^7$~yrs at  $\dot{M}_{-8}=1$ could spin-up the SS to frequency $f \sim 1$kHz,
at a nearly constant temperature (Zdunik et al.  2002) (the vertical
S-track in Fig.\ ~\ref{fig:fT}). Along the S-track, for $m_{\rm
s}=200$MeV, a star is stable with respect to the r-mode instability
because of its huge bulk viscosity. It can be accelerated to high
rotational frequency reaching the critical frequency for viscosity driven
instability (short-long dashed line in Fig.\ ~\ref{fig:fT}).  After
crossing $f=f_{\rm crit}=(0.9-1.2)\;$kHz (the precise value depending
on the SS mass and the detailed form of the SQM EOS) the SS is stable
to the bar mode formation since the characteristic growth time-scale of
viscosity driven instability $\tau_{\zeta}\sim $ a month $>>$ than tens
of seconds - the gravitational wave damping time.  In the case of
$m_{\rm s}=200$MeV a strange star can be accelerated to very high
rotational frequencies until it reaches one of the upper limits (low-T
r-mode instability region, dynamical instability limit, mass-shedding limit).

Let us stress the importance of the value of $m_{\rm s}$ for such an
astrophysical scenario.  For high mass of strange quark $> 200$~MeV a
star can be accelerated to very high frequencies without emitting
gravitational waves. If on the contrary the strange quark mass is 
$m_{\rm s}\le 100$~MeV the low-T r-mode instability window (above the dotted
line in Fig. ~\ref{fig:fT}), it  can be reached at a rotation rate of 0.4-0.6
kHz.  The rotating SS is expected to be in a quasi-stationary state,
in which the angular momentum losses via gravitational radiation are
balanced by the accretion torque. In this steady state an accreting SS
is an emitter of a persistent gravitational wave signal.

\subsubsection{The role of the magnetic field}
Our results could be changed if we take the magnetic field into account.
The presence of a strong magnetic field in SQM could also contribute
to the angular momentum transport and therefore influence 
different kinds of gravitational instabilities. The electrical
conductivity of the SQM is huge. Using the formula derived by
Heiselberg \& Pethick (1993) we get 
$\sigma \simeq (0.1/\alpha)^{5/3}\;(\rho/\rho_0)^{ 8/9}\; {\rm T}_9^{-5/3}\; 10^{21}~{\rm
s^{-1}}$. Therefore, the magnetic field is frozen in SQM which combined
with the presence of rapid and differential rotation could lead to
additional viscous mechanisms due to the magnetorotational
instability. 

Although a high conductivity is a key element of compact stars
(neutron stars, strange stars) physics there have been only a few
discussions of the effect of magnetic fields on triaxial instabilities
(Bonazzola et al. 1996, Spruit 1999, Rezzolla et al. 2000, Ho \& Lai
2000).  It was suggested by Bonazzola, Frieben \& Gourgoulhon (1996) that
the magnetic field can prevent the Dedekind-like mode. Assuming
infinite conductivity, shear of the fluid from differential angular
velocity generates some toroidal magnetic field in addition to the
poloidal one (Bonazzola \& Marck 1994). The excess of kinetic energy
contained in differential rotation with respect to rigid rotation can
be efficiently converted into magnetic energy, enforcing rigid
rotation.

Ho \& Lai (2000) have studied the combined effects of magnetic braking
and gravitational radiation generated by r-mode instability on the
spin evolution of young rapidly rotating neutron stars. They suggested
that a strong magnetic field can drive the instability just like
gravitational radiation reaction. The question of whether an unstable
r-mode leads to differential rotation, and whether the mode can be
prevented from growing by the magnetic field, was discussed by
Rezzolla, Lamb and Shapiro (2000). They studied how  so-called
Stokes drift affects the magnetic field of the star, and what the
backreaction on the mode may be.  It was estimated that the
instability could operate in young neutron stars and recycled ones
provided that they spin fast enough.

\section{Summary}
We have performed numerical investigations of the Jacobi-like bar mode
secular instability of rotating strange quark stars described by the MIT
bag model. In particular, we have computed the domain of the parameter
space of stationary rigidly rotating configurations in which 
triaxial instability develops.  It has been found that, contrary to
neutron stars, the triaxial instability can set in quite far towards the
mass-shed limit and is allowed for small mass stars as well if
the viscosity is high enough to damp out any deviation from rigid
rotation.  The lower mass is the lower rotational frequency at
which Jacobi-like instability can develop.  The viscosity driven
instability does not set an upper limit on the spin rate (but on
$T/|W|$ or oblateness) of strange stars - the configuration marginally
stable against non-axisymmetric perturbation can rotate with a frequency
higher than mass-shedding configurations (the maximal rotating
configuration is not a Keplerian one for SS). A normal and low mass
supramassive strange star gaining angular momentum always slows down
before reaching the Keplerian limit. For a high-mass supramassive
strange star sequence, the mass-shedding configuration is the one with
the lowest rotational frequency in the sequence. The instability, if
relevant, can impose a strong constrain on the lower limit of the frequency
at the innermost stable circular orbit around rapidly rotating strange
stars. The results are robust for all linear self-bound equations of
state.  The critical rotation frequency for viscosity driven
instability in the case of strange stars described by the MIT bag model is
high, $0.8-1.2 {\rm\ kHz}$ (depending on the stellar mass and the
detailed form of the SQM EOS).

 We have discussed astrophysical scenarios when triaxial instabilities
(r-mode and viscosity driven instability) could be relevant in strange
stars described by the standard MIT bag model of normal quark matter
(any kind of superconductivity being excluded). We have shown, taking
into account actual values of shear viscosity, that viscosity driven
instability cannot develop in any astrophysicaly relevant temperature
windows.

 Whether a strange star can be a good emitter of gravitational waves
  or not depends on CFS instability.  In our discussion we focused on
  the most promising CFS instability (r-mode). The spin
  evolution of strange stars and emission of gravitational waves
  strongly depend on the mass of the strange quark (caused by the strong
  dependence of bulk viscosity on the mass of a strange quark). If we
  take the strange quark mass to be $m_{\rm s}c^2 \sim 200\ {\rm MeV}$
  or higher (consistent with renormalization and particle data tables)
  then r-mode instability is relevant only in a narrow temperature
  window and seems to be unimportant in Low Mass X-Ray Binaries. In
  this case an old star can be spun up in LMXBs to very high rotation
  rates and a low frequency at the marginally stable orbit can be
  obtained ($\sim 1$kHz).

 If on the contrary the strange quark mass is low, $m_{\rm s}c^2 \sim
 100 {\rm\ MeV}$, then a star accelerated in LMXBs to a frequency of
 $0.4-0.6\ $kHz can be a good emitter of persistent gravitational
 waves (see Andersson, Jones, Kokkotas 2002 for a discussion) that
 ought to be detectable with large-scale interferometers.  In this
 case the r-mode instability can impose strong constraints on the
 lower limits on rotational period and then on the maximal orbital
 frequency in LMXBs.  The lowest frequency at the marginally stable
 orbit could be obtained for slowly-rotating massive strange stars.

 A young strange star would not be spun down significantly by the
  r-modes during its first few months of existence (because of its very
  rapid cooling rate) until the star has cooled to a temperature of a
  few times $ 10^7- 10^8$~K.

  The astrophysical scenario will change if values of viscosity change
  or/and different physical mechanisms e.g. magnetic field will be
  taken into account.

\begin{acknowledgements}
We are grateful to N. Stergioulas, N. Andersson, L. Villain,
 S. Bonazzola, D. Lai, J. Friedman and K. Kokkotas for helpful
 discussions. This work has been funded by the following grants: KBN
 grants 5P03D.017.21 and 5P03D.020.20; the Greek-Polish Joint Research
 and Technology Programme EPAN-M.43/2013555 and the EU Programme
 ``Improving the Human Research Potential and the Socio-Economic
 Knowledge Base'' (Research Training Network Contract
 HPRN-CT-2000-00137).

\end{acknowledgements}

\end{document}